\documentclass[bm,aps,
amsfonts,amssymb,
preprint,
nofootinbib]{revtex4}

\usepackage[dvipdfmx]{graphicx}
\usepackage{natbib}
\usepackage{amsmath}

\usepackage{color}

\font\mybb=msbm10 at 12pt

\font\mybbsmall=msbm10 at 10pt

\def\bb#1{\hbox{\mybb#1}}

\def\bbsmall#1{\hbox{\mybbsmall#1}}

\def\ZZ {\bb{Z}}

\def\ZZsmall {\bbsmall{Z}}

\def\PP {\bb{P}}

\def\PPsmall {\bbsmall{P}}

\def\AA{{\bf A}}
\def\BB{{\bf B}}
\def\CC{{\bf C}}

\newcommand\beqa{\begin{eqnarray}}
\newcommand\eeqa{\end{eqnarray}}
\newcommand\n{\nonumber\\}

\begin{document}

{~}

\title{More on a dessin on the base: Kodaira exceptional fibers \\and  
mutually (non-)local branes
}
\author{Shin Fukuchi\footnote[1]{E-mail:fshin@post.kek.jp},
Naoto Kan\footnote[2]{E-mail:naotok@post.kek.jp},
\\
Rinto Kuramochi\footnote[3]{E-mail:rinto@post.kek.jp},
Shun'ya Mizoguchi\footnote[4]{E-mail:mizoguch@post.kek.jp}
and Hitomi Tashiro\footnote[5]{E-mail:tashiro@post.kek.jp}}


\affiliation{\footnotemark[4]Theory Center, 
Institute of Particle and Nuclear Studies,
KEK\\Tsukuba, Ibaraki, 305-0801, Japan 
}

\affiliation{\footnotemark[1]\footnotemark[2]\footnotemark[3]\footnotemark[4]\footnotemark[5]SOKENDAI (The Graduate University for Advanced Studies)\\
Tsukuba, Ibaraki, 305-0801, Japan 
}

\begin{abstract} 
A ``dessin d'enfant'' is a graph 
embedded on a two-dimensional oriented surface  
named by Grothendieck. Recently 
we have developed a new way to keep track 
of non-localness among 7-branes in F-theory on an elliptic 
fibration over $\PPsmall^1$ by drawing a triangulated 
``dessin'' on the base.
To further demonstrate the usefulness of this method,
we provide three examples of its use. 
We first consider a deformation of the $I_0^*$ Kodaira 
fiber. With a dessin,  
we can immediately find out which pairs of 7-branes 
are (non-)local and compute their monodromies.  
We next identify the paths of 
string(-junction)s on the dessin by solving the mass 
geodesic equation.  By numerically computing their total masses,  
we find that the Hanany-Witten effect 
has not occurred in this example.
Finally, 
we consider the orientifold limit 
in the spectral cover/Higgs bundle approach. 
We observe the characteristic configuration 
presenting the cluster sub-structure of an O-plane 
found previously.

\end{abstract}

\preprint{KEK-TH-2172}
\date{December 5, 2019}

\maketitle

\newpage
\section{Introduction}
A ``dessin d'enfant'' is a graph 
embedded on a two-dimensional oriented surface  
named by Grothendieck \cite{Grothendieck,LandoZvonkin}. Recently,  
we have shown that such a drawing on a base $\PP^1$ of 
an elliptic fibration, along with a triangulation, is a convenient 
tool for computing the monodromies of the elliptic fiber \cite{dessinonthebase}.  
In this set-up, we introduce two kinds of new codimension-one 
objects defined by the zero loci of the coefficient functions 
$f(z)$ and $g(z)$ of the Weierstrass equation. They correspond 
to the two kinds of nodes of a dessin d'enfant. We draw lines 
on the $\PP^1$ base at the preimages of the modular $J$-function for 
$-\infty<J(\tau(z))<0$, $0<J(\tau(z))<1$ and $1<J(\tau(z))<\infty$, 
which we call $T$-wall, 
$S$-wall and $T'$-wall, respectively. They amount to constitute 
a triangulated dessin associated with a special Belyi function.  
We have shown that the monodromies around 7-branes along 
an arbitrary path can be very easily computed once such a 
dessin is drawn. We have also shown that the cells encircled by 
the $S$-walls are connected regions of weak coupling, with which 
we can identify which pairs of 7-branes are mutually local or 
non-local.

In this paper, to further demonstrate the usefulness of this method,  
we consider the following three examples: 

\noindent
(i) We first consider a certain deformation of a $I_0^*$ Kodaira fiber  
and examine which pairs of 7-branes 
are (non-)local using the dessin.
We compute their monodromies 
and compare them with the conventional {\bf A}{\bf B}{\bf C} 
description of 7-branes.

\noindent
(ii) We next identify the paths of 
string(-junction)s on the dessin by solving the mass 
geodesic equation.  By numerically computing their total masses,  
we find that the Hanany-Witten effect 
has not occurred in this example.

\noindent
(iii) Finally, we consider 
the orientifold limit 
in the spectral cover/Higgs bundle approach. 
We observe the characteristic configuration 
presenting the cluster sub-structure of an O-plane, 
which was found previously in \cite{dessinonthebase}.

The plan of this paper is as follows. 
In the next section, we give a review of the dessin on the base 
developed in \cite{dessinonthebase}. 
In section 3, we consider an example of a 
deformation of a $I_0^*$ Kodaira fiber. 
In section 4, we solve the mass geodesic equation 
and compute the total masses of string(-junction)s numerically.
In section 5, we use this method in an example of 
the spectral cover/Higgs bundle approach.
The last section is devoted to conclusions.

\section{A brief review of the dessin on the base}
Let us summarize how a ``dessin on the base'' can be 
drawn on $\PP^1$ \cite{dessinonthebase}. 
We consider F-theory on an elliptic fibration over $\PP^1$ with a section,
given by a Weierstrass equation 
\beqa
y^2&=&x^3+f(z)x+ g(z),
\eeqa
where $z$ is the coordinate of an affine patch. 
On this $z$-plane, we first plot points of the zero loci 
of $f(z)$ and $g(z)$. Just like 7-branes, 
they define codimension-one objects,  
which we call ``elliptic point planes''.\footnote{This 
name comes from the fact that the points 
$\tau=e^{\frac{2\pi i}3}$ and $i$ in the fundamental 
region are called ``elliptic points'', which means that 
their isotropy groups consist of elliptic elements of
$SL(2,\ZZsmall)$.} They correspond to two kinds 
of critical points in the ``dessin d'enfant'' construction 
of Grothendieck \cite{Grothendieck,LandoZvonkin}.

We next draw lines at the preimages of 
$-\infty<J(\tau(z))<0$, $0<J(\tau(z))<1$ and 
$1<J(\tau(z))<\infty$, respectively, which we call 
$T$-wall, $S$-wall and $T'$-wall. 
Here $J(\tau)$ is the modular $J$-function.  
 Together with 
 the discriminant loci $\Delta=4f^3+27g^2=0$, 
they constitute a dessin drawn on $\PP^1$ 
with a canonical triangulation. Concrete examples can be 
found in the following sections.

What is new in our construction is that we draw 
the three kinds of walls with different lines; specifically 
we draw a $T$-wall with a (light) green line, 
an $S$-wall with a (dark) blue line and 
a $T'$-wall with a dashed green line.  
With this we are able to read off monodoromies 
along arbitrary paths in an amazingly simple way.

The reason why we can do this in this set-up is the following. 
The modular parameter $\tau(z)$ of an elliptic fiber over 
$z\in\PP^1$ is determined by the inverse of the $J$-function:
\beqa
J(\tau(z))&=&\frac{4 f(z)^3}{4 f(z)^3+27 g(z)^2}, \label{Jequation}
\eeqa 
which defines a special Belyi function, 
in which the ramification index of the $J=0$ critical point is 
always three, and the ramification index of the $J=1$ critical 
point is always two. Since, on the other hand, $J(\tau)$ 
also behaves like $J(\tau)\sim O((\tau-e^{\frac{2\pi i}3})^3)$ 
near $\tau\sim e^{\frac{2\pi i}3}$ 
and $J(\tau)\sim 1+O((\tau-i)^2)$ near $\tau\sim i$, 
the equation (\ref{Jequation}) induces a local 
homeomorphism between the $z$-plane and the upper-half plane. 
In other words, a dessin can provides us with a precise chart on the base 
showing the corresponding position on the upper-half plane 
\footnote{The idea of computing monodromies 
by keeping track of the value of the $J$-function was developted 
by Tani \cite{Tani}.
}

Let us now spell out how our new method of reading off the 
monodromies works. Following the notation used in 
\cite{dessinonthebase}, we denote a $T$-wall as {\bf G} (for {\bf G}reen),
an $S$-wall as {\bf B} (for {\bf B}lue) 
 and a $T'$-wall as {\bf dG} (for {\bf d}ashed {\bf G}reen).
 To compute a monodromy along a given path, 
one first lists the walls crossed by the path in the order that 
they appear along the path. Then, if the starting point of the 
path is in a shaded cell region\footnote{A shaded cell region is 
referred to 
a cell 
region with ${\rm Im} J> 0$, whereas an unshaded cell region is 
one with ${\rm Im} J < 0$ \cite{dessinonthebase}.}, one assigns for 
each two crossings a particular $SL(2,\ZZ)$ matrix 
according to the following rule, and subsequently multiplies 
it from the right inductively:
 \beqa
 \rightarrow {\bf dG} \rightarrow  {\bf G} \rightarrow &=&T, \n 
  \rightarrow {\bf G} \rightarrow  {\bf dG} \rightarrow &=&T^{-1}, \n 
  \rightarrow {\bf dG} \rightarrow  {\bf B} \rightarrow &=&  ~\rightarrow {\bf B} \rightarrow  {\bf dG} \rightarrow ~=~S, \n 
  \rightarrow~ {\bf B}~ \rightarrow  {\bf G} \rightarrow &=&ST, \n 
  \rightarrow ~{\bf G}~ \rightarrow  {\bf B} \rightarrow &=&T^{-1}S.
  \label{rule}
\eeqa
Also, if the the starting point of the 
path is in a unshaded cell region, the rule is 
 \beqa
 \rightarrow {\bf dG} \rightarrow  {\bf G} \rightarrow &=&T^{-1}, \n 
  \rightarrow {\bf G} \rightarrow  {\bf dG} \rightarrow &=&T, \n 
  \rightarrow {\bf dG} \rightarrow  {\bf B} \rightarrow &=&  ~\rightarrow {\bf B} \rightarrow  {\bf dG} \rightarrow ~=~S, \n 
  \rightarrow~ {\bf B}~ \rightarrow  {\bf G} \rightarrow &=&ST^{-1}, \n 
  \rightarrow ~{\bf G}~ \rightarrow  {\bf B} \rightarrow &=&TS.
  \label{ruleforunshaded}
\eeqa
For more detail of this method, see \cite{dessinonthebase}.

\section{(non-)local 7-brane pairs in a 
deformation of a $I_0^*$ Kodaira fiber on a dessin}

Let us consider a deformation of a $I_0^*$ Kodaira fiber.
The coefficient functions of the Weierstrass equation is 
given by
\beqa
f(z)&=&(z+1)(z-1),\\
g(z)&=&z(z^2+1).\nonumber
\eeqa
The discriminant is 
\beqa
\Delta&=&4f(z)^3+27g(z)^2\n
&=&31 z^6+42 z^4+39 z^2-4.
\eeqa

\begin{figure}[h]%
\centerline{
\includegraphics[width=0.5\textwidth]{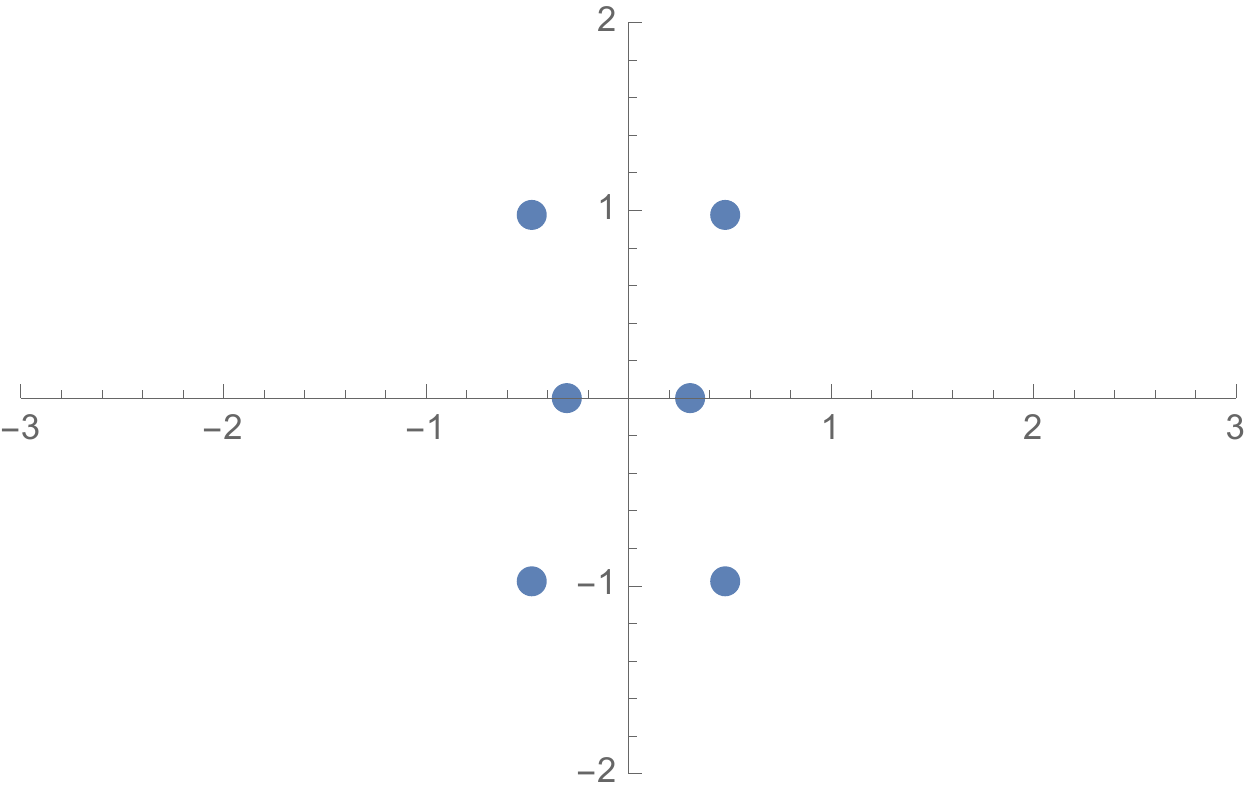}}
\caption{\label{discrloci}
}
\end{figure}

Six 7-branes take positions at the zero loci of the discriminant,
which are shown in FIG.\ref{discrloci}.
If they gather altogether to a single point, they constitute 
a $I_0^*$ fiber. This is the resolution of one of the four singularities of the 
eight-dimensional orientifold compactificaiton \cite{Senorientifold}. 
The deformation is also the total space of 
a Seiberg-Witten curve of ${\cal N}=2$ $SU(2)$ SUSY 
gauge theory with four flavors \cite{SW}.

If we are given only the positions of the 7-branes alone, 
we cannot tell which two of them are local or non-local. 
The canonical presentation of the $I_0^*$ fiber in the
{\bf A}{\bf B}{\bf C} description \cite{GHZ,DZ,DHIZ}
is {\bf A}{\bf A}{\bf A}{\bf A}{\bf B}{\bf C}.
Which brane is {\bf B} and which brane is {\bf C}?
In fact, it turns out that it is not natural to identify these 
six 7-branes as a set of four {\bf A}-branes, one {\bf B}-brane 
and one {\bf C}-branes.

\begin{figure}[h]%
\centerline{
\includegraphics[width=0.55\textwidth]{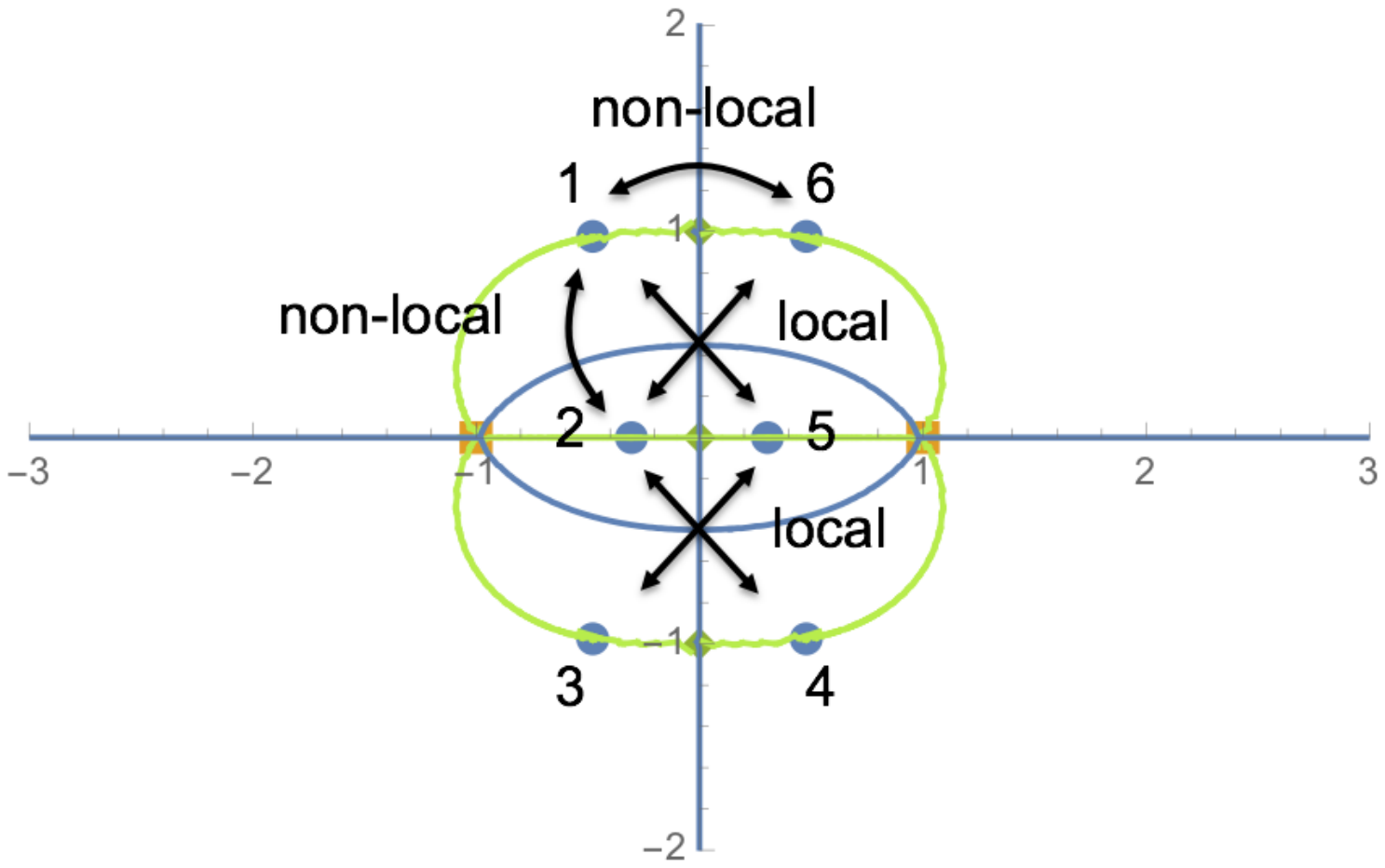}
}
\caption{\label{deformedI0stardessin}}
\end{figure}

\begin{figure}[h]%
\centerline{
\includegraphics[width=0.5\textwidth]{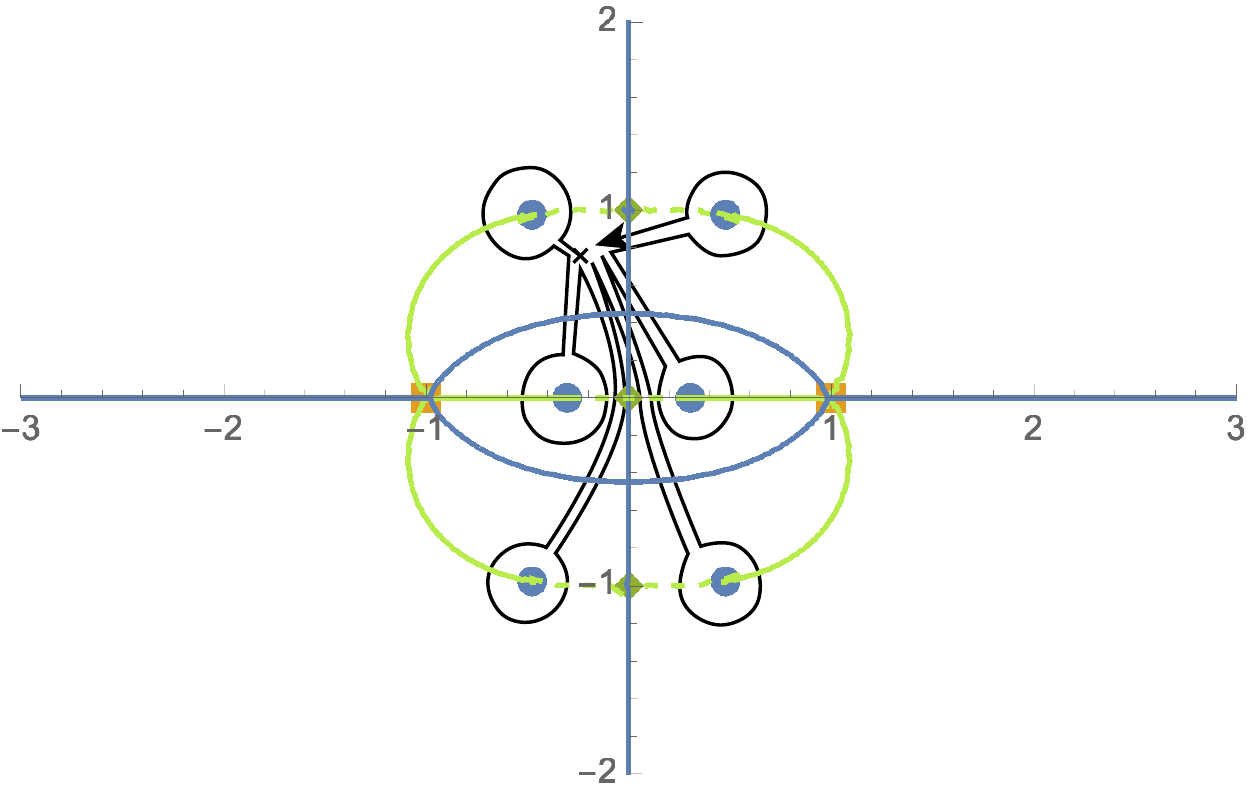}
}
\caption{\label{deformedI0stardessinwithcontour}}
\end{figure}

The dessin for this configuration is shown in FIG.\ref{deformedI0stardessin}.
As we mentioned in Introduction, (light) green lines are the $T$-walls, 
(dark) blue lines are the $S$-walls and dashed green lines are the $T'$-walls. 
The locations of the $f$-planes (the loci of $f=0$) are denoted by 
squares, those of the $g$-planes (the loci of $g=0$) are shown 
by $45^\circ$-rotated squares and the 7-branes are represented 
by circles.   
By this figure one can immediately see that the 7-branes {\bf 1},
{\bf 3} and {\bf 5} are 
mutually local, and the ones {\bf 2}, {\bf 4} and {\bf 6} 
are mutually local, and 
one in the first set and another in the second set are mutually nonlocal.
This is because, for instance, the straight path connecting the brane {\bf 1} 
and the brane {\bf 5} goes through only $S$-walls twice, which means that 
the monodromy is trivial. 

The important point is that the notion of whether a 
particular pair of two 7-branes are local or non-local 
is fundamental-group dependent. For instance, one can 
also connect the branes {\bf 1} and {\bf 5} through a path 
passing the left side of the brane {\bf 2}. In this case the monodromy 
is nontrivial, and the two 7-branes are non-local along this path. 
This point is further elaborated in the next section.

In fact, if we choose the loop of the monodromy to be as 
shown in FIG.\ref{deformedI0stardessinwithcontour}, 
the monodromy matrices are
\beqa
\AA_1{\bf N}_2\AA_3{\bf N}_4\AA_5{\bf N}_6,
\eeqa 
where we have added the labels of the branes 
 shown in FIG.\ref{deformedI0stardessin}  as subscripts. 
The definitions of the monodromy matrices are  
\beqa
{\bf A}&=&M_{1,0}~=~\left(
\begin{array}{rc}
1~~&1\\ 0~~&1
\end{array}
\right)~=~T,\n
{\bf N}&=&M_{0,1}~=~\left(
\begin{array}{rc}
1~~&0\\ -1~~&1
\end{array}
\right)~=~STS,\n
{\bf B}&=&M_{1,1}~=~\left(
\begin{array}{rc}
2~~&1\\ -1~~&0
\end{array}
\right)~=~T^{-2}S, \n
{\bf C}&=&M_{1,-1}~=~\left(
\begin{array}{rc}
0~~&1\\ -1~~&2
\end{array}
\right)~=~ST^{-2},
\eeqa
where
\beqa
T=\left(
\begin{array}{rc}
1~~&1\\ 0~~&1
\end{array}
\right),~~~
S=\left(
\begin{array}{rc}
0~~&-1\\ 1~~&0
\end{array}
\right).
\eeqa
The equalities hold as elements  
of $PSL(2,\ZZ)$ (that is, up to a factor of $-1$). 
Note that as a product of matrices 
$\AA{\bf N}\AA{\bf N}\AA{\bf N}=\AA\AA\AA\AA\BB\CC$.
%
\begin{figure}[b]%
\centerline{
\includegraphics[width=1.0\textwidth]{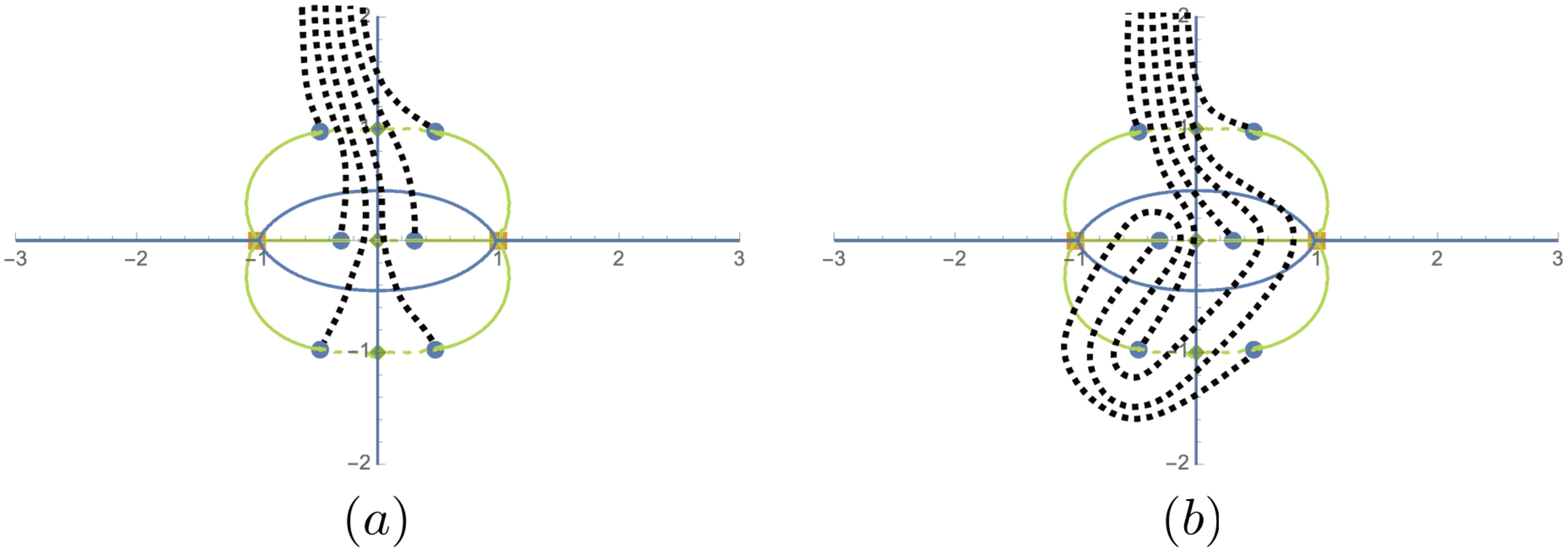}
}
\caption{\label{cuts}}
\end{figure}

Specifying a loop of the monodromy amounts to 
choosing the branch cuts extending from the 7-branes.
The loop in FIG.\ref{deformedI0stardessinwithcontour}, 
with which the monodromy is $\AA{\bf N}\AA{\bf N}\AA{\bf N}$, 
the corresponding configuration of the branch cuts 
is shown in FIG.\ref{cuts}(a). This is the ``canonical''
representation. On the other hand, one can work out a  
configuration of the branch cuts that yield the monodromy 
$\AA\AA\AA\AA\BB\CC$. The result is shown in 
FIG.\ref{cuts}(b).\footnote{This is not a unique set of branch cuts 
that give $\AA\AA\AA\AA\BB\CC$, but there is no simpler one 
than this. 
} 

As we can see, it is complicated. 
Although the 7-brane {\bf 4} is also seen to be ${\bf A}$ in 
this choice of the cuts (and hence the choice of the loop), 
an $(1,0)$-string extended from it can end on, say, the brane {\bf 3} 
only when the string takes a long detoured route around the brane {\bf 2} along the cut.  

In fact, if we choose the fundamental region of the 
starting point as the standard one, we get a monodromy 
$\AA\AA\AA\AA{\bf M}_{2,1}{\bf N}$ instead of 
$\AA\AA\AA\AA\BB\CC$; they are related by a conjugation 
of $T(=\AA)$: $\BB=T{\bf M}_{2,1}T^{-1}$, 
$\CC=T{\bf N}T^{-1}$.
Therefore, in order to have $\AA\AA\AA\AA\BB\CC$, 
we need either to start from the fundamental region 
next to the standard one, or, starting from the standard 
fundamental region,  go once around the brane {\bf 1} 
anti-clockwise before computing the monodromy and clockwise 
again after going back and forth among the branes.  

One can also work out the relations between the string(-junction)s 
extending from the branes $\AA{\bf N}\AA{\bf N}\AA{\bf N}$ 
and those from $\AA\AA\AA\AA\BB\CC$. 
For example, writing the ordered 
set of branes as
\beqa
\AA_1{\bf N}_2\AA_3{\bf N}_4\AA_5{\bf N}_6
&=&\AA'_1\AA'_3\AA'_5\AA'_4\BB_2\CC_6,
\eeqa
the string $a'_3-a'_4$
on the right hand side
(in the notation of \cite{DZ}; primes are put 
to distinguish them from the {\bf A}-branes on the
left hand side)
is expressed as $n_4-n_2$ on the left hand side. 
This can be understood as a process of two 
subsequent Hanany-Witten effects. Using the standard 
7-brane technology \cite{DZ},
the full set of independent relations are found to be:
\beqa
a_1-a_3&=&a'_1-a'_3,\n
a_3-a_5&=&a'_3-a'_5,\n
n_2-n_4&=&a'_4-a'_3,\n
n_4-n_6&=&-a'_5-a'_4+b_2+c_6.
\eeqa

\section{Hanany-Witten effect and mass geodesics of 
string(-junction)s on a dessin}

\begin{figure}[h]%
\centerline{
\includegraphics[width=1.0\textwidth]{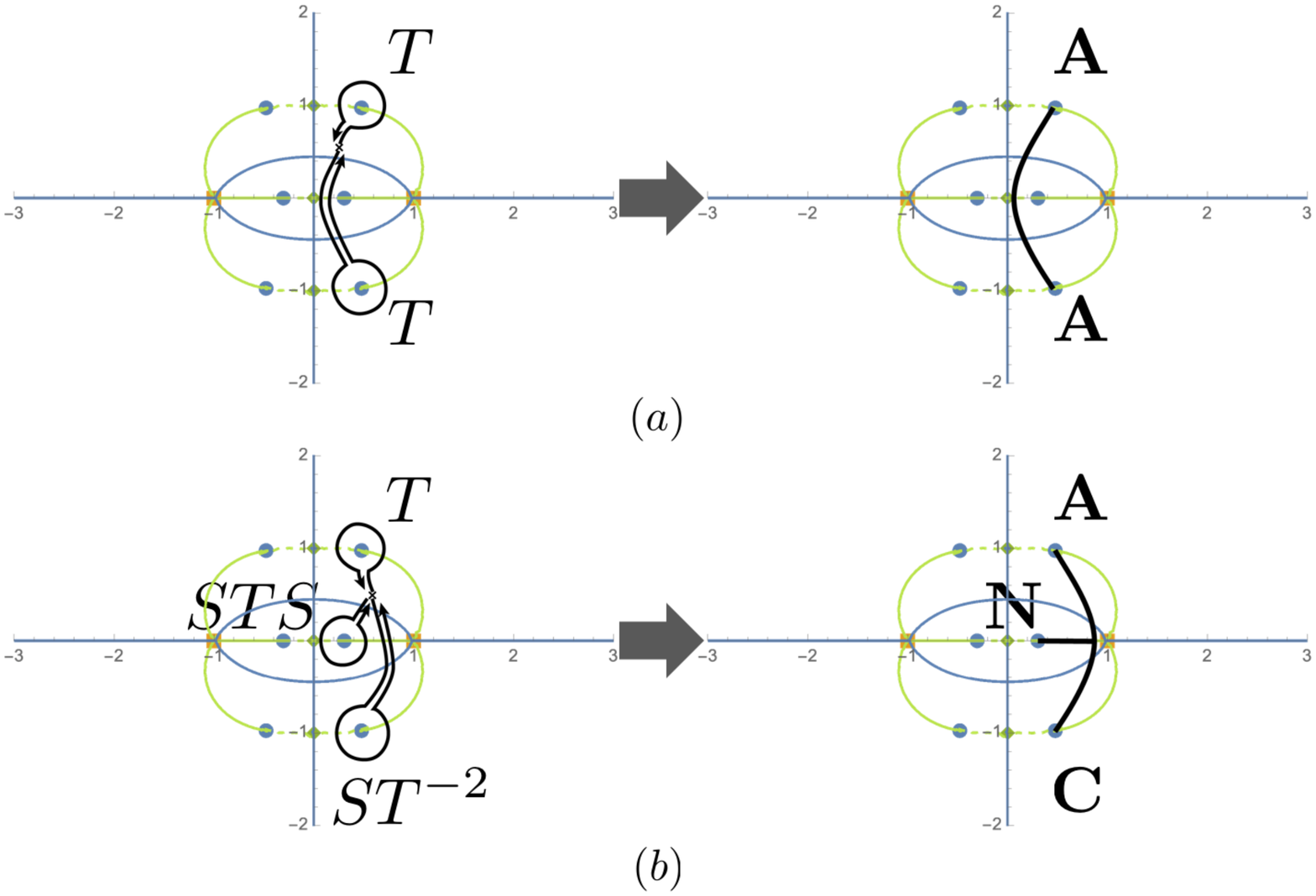}
}
\caption{\label{AA_vs_ANC}}
\end{figure}

Next we consider the Hanany-Witten effect on a dessin.
Let us consider a string or a string junction connecting 
the branes {\bf 4} and {\bf 6} in the example in previous section. 
As we saw there, these two branes are seen to be 
mutually local if their $(p,q)$-charges are measured 
(that is, the monodoromies are computed) from the 
left side of the brane {\bf 5} (FIG.\ref{AA_vs_ANC}(a)). 
On the other hand, if we take the starting point of 
the loop near the brane {\bf 4} and measure the 
$(p,q)$-charge of {\bf 6} along the path passing on the right 
side of the brane {\bf 5}, the monodoromy around {\bf 4} 
is $T$ while that around {\bf 6} is $ST^{-2}$, implying 
that {\bf 4} is {\bf A} and {\bf 6} is {\bf C}.
Thus we cannot join them with an open string directly 
through this side because the open 
string stretched on this side ``observes'' the branes {\bf 4} and {\bf 6} 
as  {\bf A} and {\bf C}.
In this case, however, the brane {\bf 5} is seen to be {\bf N}, so 
we can consider a string junction with an extra $(0,-1)$-string 
extending from the brane {\bf 5}(FIG.\ref{AA_vs_ANC}(b)).
Note that the monodromy matrices only change by a  
simultaneous $SL(2,\ZZ)$ conjugation  
 if the starting point is chosen at some different place 
 as far as the fundamental group of the loop does not
change.

Thus there are two ways of connecting the branes {\bf 4} 
and {\bf 6}: One by connecting them by an open string passing on 
the left side of  {\bf 5}, and one by connecting them 
by a string junction with a junction point being on the right 
side of {\bf 5}. 
(There are certainly infinitely many more ways, 
but we do not consider them here since they will be heavier than the  
two cases we consider here.) 

\begin{figure}[h]%

\centerline{
\includegraphics[width=0.8\textwidth]{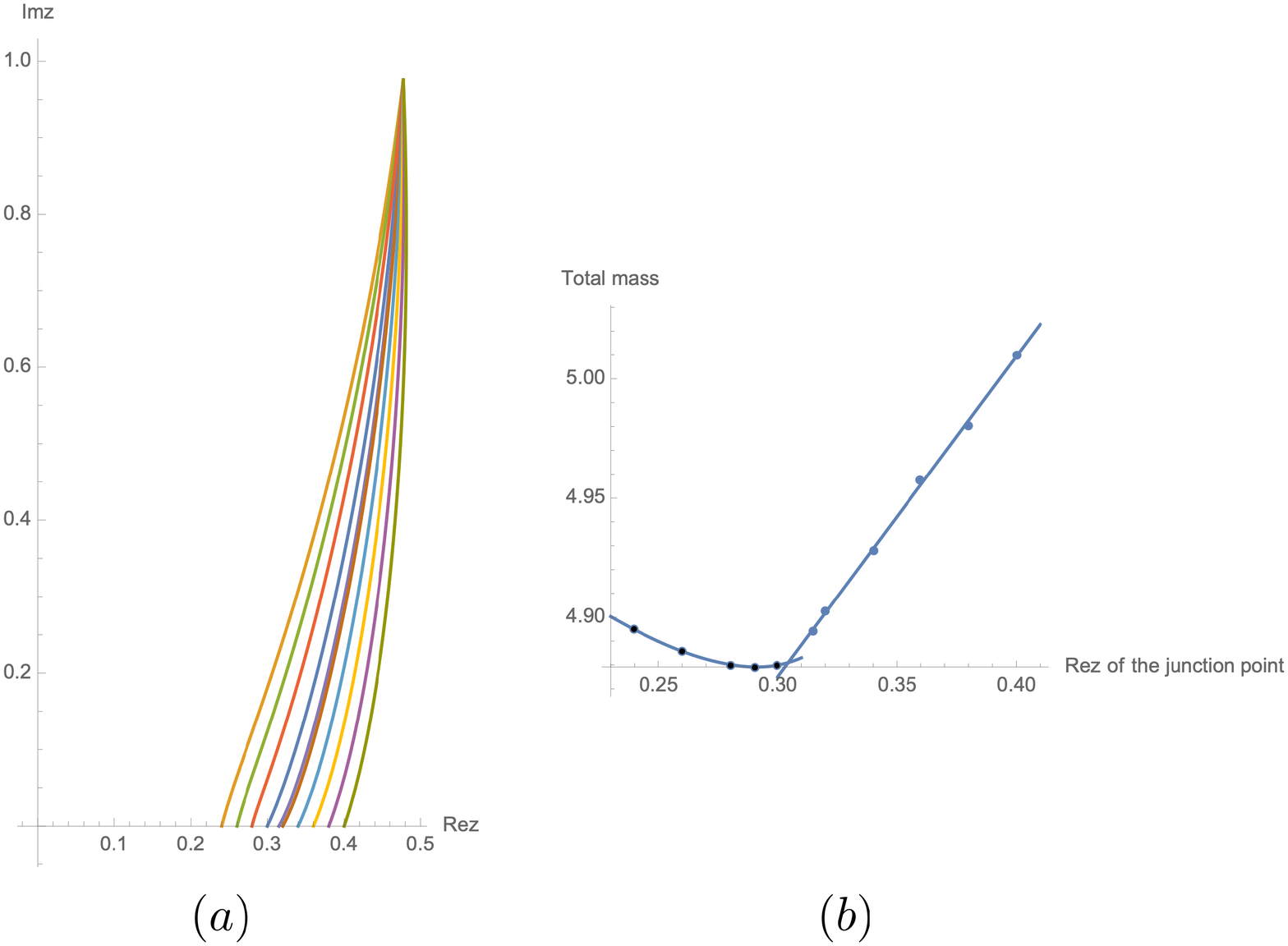}
}
\caption{\label{massgeodesicstringvsstringjunction}}
\end{figure}

Let us examine which gives the lighter state.  
In the eight-dimensional compactification of F-theory, 
where the configuration varies depending only on $z$, the metric 
of an infinitesimal line element is given by \cite{cosmicstring}
\beqa
ds^2&=&-dt^2+\sum_{i=1}^7(dx^i)^2 + e^{\varphi(z,\bar z)}dz d\bar z,
\label{ds2}\\
e^{\varphi(z,\bar z)}&=&\frac{\tau(z)-\bar\tau(\bar z)}{2i} \eta^2(\tau(z))
\bar\eta^2(\bar\tau(\bar z))
\prod_{i=1}^N (z-z_i)^{-\frac1{12}} (\bar z-\bar z_i)^{-\frac1{12}}.
\label{etophi}
\eeqa
All the other fields are set to zero.
$N$ is the number of 7-branes that can be seen 
on this affine patch of $\PP^1$. We define the effective ``mass metric''
by multiplying the tension square of a $(p,q)$-string 
\cite{Schwarz9508143} as
\beqa
ds_{mass}^2&=&T_{p,q}^2 ds^2,\\
T_{p,q}&=&\frac1{\sqrt{\rm{Im} \tau}}|p+ q\tau|.
\label{tension}
\eeqa
The geodesic length of this effective metric gives the mass of 
a string or a string junction. 
For an actual computation, it is convenient to use the fact that 
\cite{Sen9608005}
\beqa
\left|ds_{mass} \right| &=& |p da+ q da_D|
\label{pda+qdaD}
\eeqa
up to a constant factor, where $a$ is the Coulomb branch 
parameter of the gauge theory and $a_D$ is its dual \cite{SW},
while the gauge invariant ``$u$'' parameter is the $z$ coordinate 
here. In the standard fundamental region, a fundamental 
($(p,q)=(1,0)$) string is the lightest of all. It can also be shown that 
\beqa
|da|&=&|\omega_1 dz|,
\eeqa
where 
$\omega_i$ $(i=1,2)$ are the periods of the curve 
of the two homology cycles $\alpha, \beta$ of the 
elliptic fiber 
\begin{eqnarray}
\int_\alpha \omega = \omega_1,~~~
\int_\beta \omega = \omega_2
\label{periods}
\end{eqnarray}
and $\omega$ is the holomorphic differential.
Thus all we need to do is consider the metric
\beqa
ds_{mass}^2&=&|\omega_1|^2dz d\bar z,
\eeqa
solve the geodesic equation, and 
compute the geodesic length numerically.
The result is shown in 
FIG.\ref{massgeodesicstringvsstringjunction}.

Since the brane configuration is symmetric 
under the reflection with respect to the real axis, we have 
numerically computed the contours of mass geodesics from 
the brane {\bf 6} to various crossing points on 
the real axis (the left panel). The position of the brane {\bf 5} 
is $z=0.304435$. If the crossing point on the real axis is to the left
of the brane {\bf 5}, the total mass of the string connecting the branes 
{\bf 4} and {\bf 6} is twice as the ``mass geodesic'' length 
connecting the brane {\bf 4} and the crossing point. 
If, on the other hand, the crossing point is to the right of 
the brane {\bf 5}, there arises another $(0,-1)$-string stretching 
from the brane {\bf 5} to the crossing point, which adds to 
the total mass. We have shown, 
in the right panel of FIG.\ref{massgeodesicstringvsstringjunction}, 
the numerical result of the total mass 
of the string or the string junction for various values of the 
crossing point.  

From this figure we can see that the string crossing at 
around $z=0.29$ is the lightest. This string is lighter 
than any string junction with any crossing (=junction) point. 
The total mass of a string junction increases linearly 
as the junction point becomes far away from the brane {\bf 5} 
because of the contribution of the third string stretched from 
the brane {\bf 5}.

The Hanany-Witten effect \cite{HananyWitten} 
is usually referred to as 
a transition of an open string into a trivalent string junction, 
from such as 
one shown FIG.\ref{AA_vs_ANC}(a) to one shown in (b). 
Since it  
may be regarded as a phenomenon in 
which a heavier string junction state 
becomes lighter than an open string state, 
we may say that, in the present brane configuration, 
this effect has not occurred for the string 
between the branes {\bf 4} and {\bf 6}.

\section{Orientifold limit in the spectral cover/Higgs bundle approach}
In F-theory, the spectral cover describes the vector bundle of 
the dual heterotic theory 
\cite{FMW,Curio,DiaconescuIonesei,DonagiWijnholt}. 
The coefficients of the  
defining equation of the spectral cover comprise 
a weighted projective space bundle, and they 
can be identified as the Casimir elements of the Lie algebra 
of the Higgs field  \cite{BHV,DonagiWijnholt}. 
Each root of the characteristic/spectral cover 
equation corresponds to a section of a rational elliptic surface 
(a $\frac12$K3 surface),
and hence a weight of the Mordell-Weyl lattice \cite{OguisoShioda},
which is typically the weight lattice of the Lie algebra of the 
Higgs.\footnote{Very recently, the construction of a $\frac12$K3 
surface is generalized to complex-three and -four dimensions 
with interesting applications to F-theory compactifications \cite{Kimura1,Kimura2}.}

In \cite{DonagiWijholtHiggsbundle}, 
the orientifold limit was considered 
in the spectral cover/Higgs bundle set-up. The Weierstrass equation is 
taken to be
\beqa
y^2 &=&x^3+ f x+ g,\n
f&=&-\frac1{48}(\texttt b_2^2 - 24 \epsilon \texttt b_4),\n
g&=&-\frac1{864}(-\texttt b_2^3 +36 \epsilon \texttt b_2 \texttt b_4 
-216 \epsilon^2  \texttt b_6).
\eeqa
The discriminant is 
\beqa
\Delta&=&-\frac14 \epsilon^2 \texttt b_2^2 (\texttt b_2 \texttt b_6-\texttt b_4^2)
+O(\epsilon^3).
\eeqa
In \cite{DonagiWijholtHiggsbundle}, the leading-order discriminant 
components are respectively identified as:
\beqa
{\mbox O7}:\texttt b_2=0,~~~\mbox{D7}:\texttt b_2 \texttt b_6-\texttt b_4^2=0.
\label{O7D7}
\eeqa
Let us confirm this by  
drawing the dissin.

\begin{figure}[h]%
\centerline{
\includegraphics[width=0.45\textwidth]{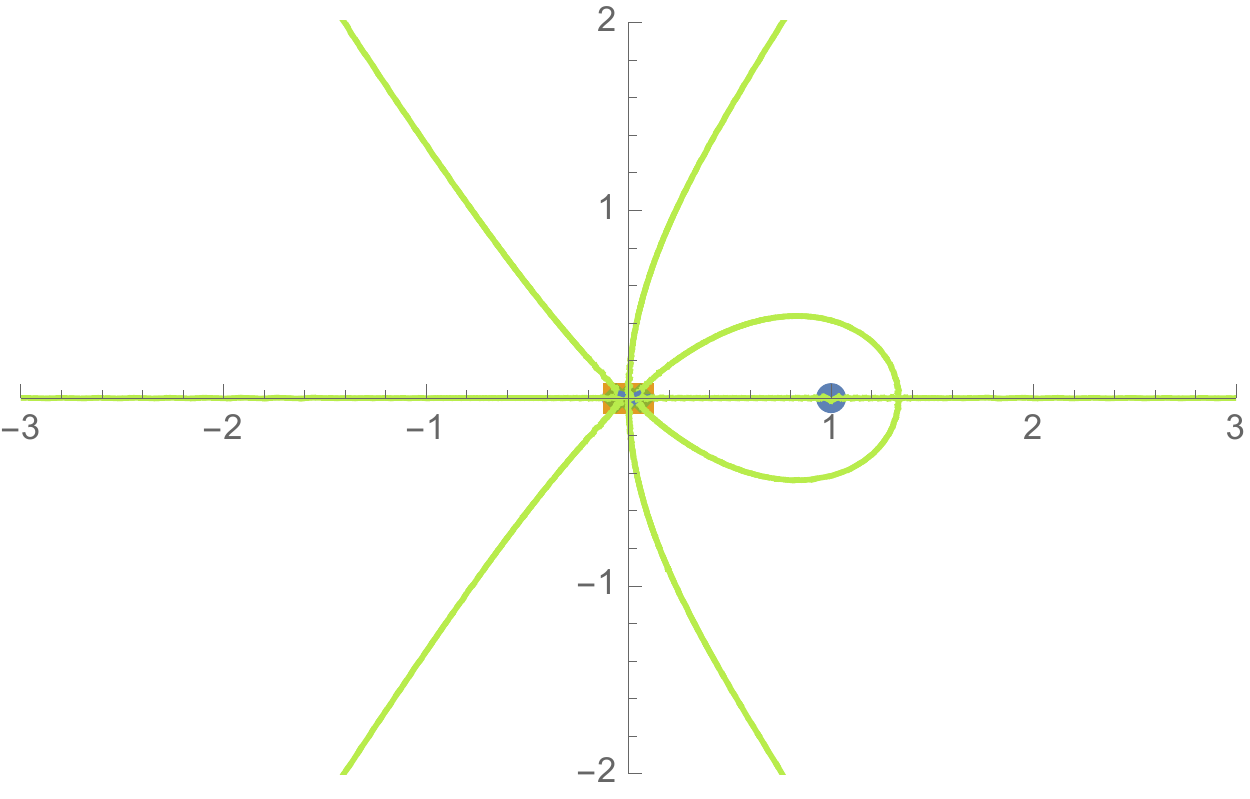}
}
\caption{\label{I0starHiggsbundle}}
\end{figure}
\begin{figure}[h]%
\centerline{
\includegraphics[width=0.45\textwidth]{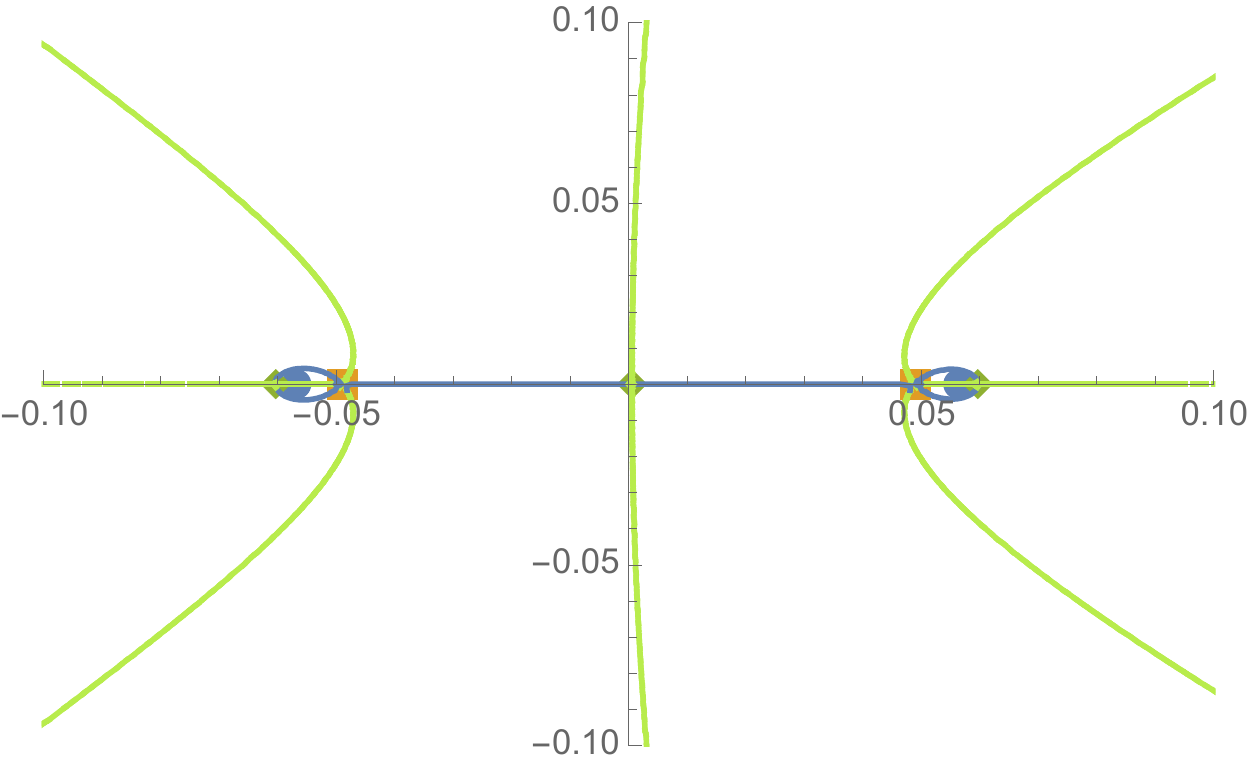}
}
\caption{\label{I0starHiggsbundlelarge}}
\end{figure}
\begin{figure}[h]%
\centerline{
\includegraphics[width=0.45\textwidth]{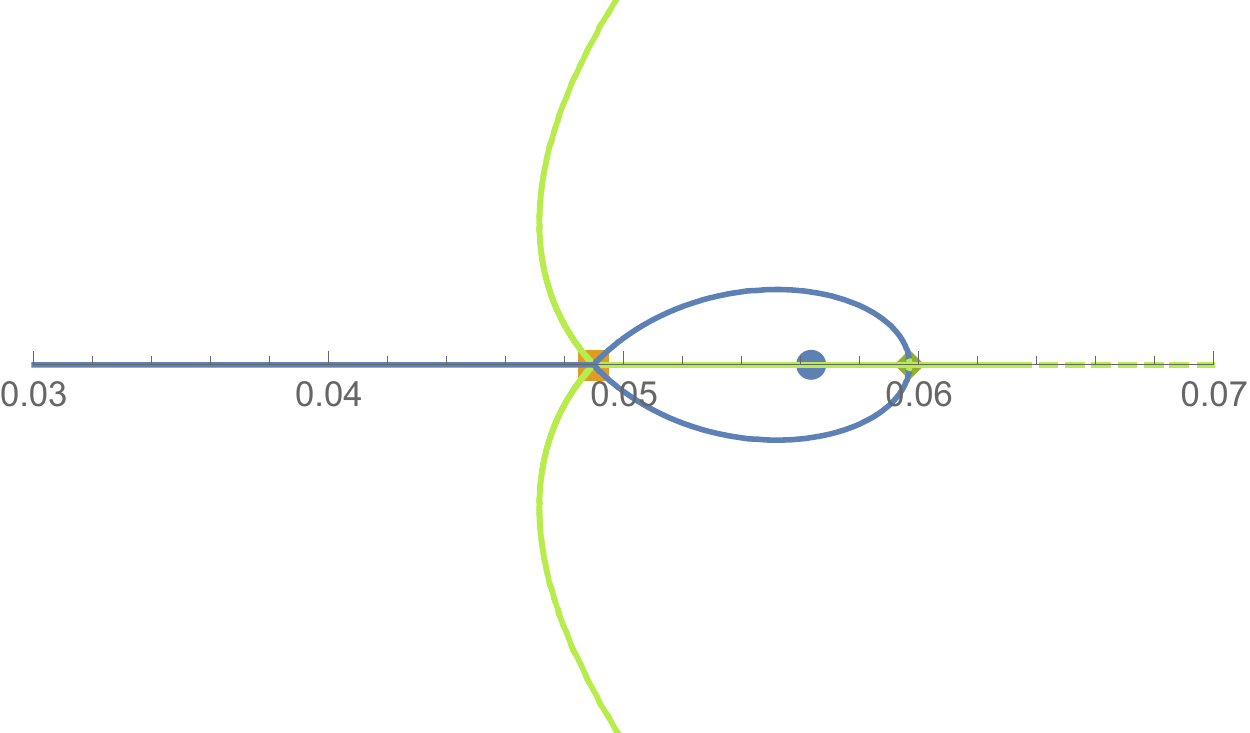}
}
\caption{\label{I0starHiggsbundlecloser}}
\end{figure}

Fig.\ref{I0starHiggsbundle} is the dessin on the base drawn for $\epsilon=0.0001$
with $\texttt b_4=\texttt b_6=1$
on the $\texttt b_2$-plane.
Since the discriminant is a cubic polynomial in $\texttt b_2$ (as 
the higher-order terms cancel), we can see three 7-branes, 
two very close to the origin and one isolated from them at 
around $\texttt b_2=1$. The former is to be identified as the O7-plane, 
whereas the latter is the D7-brane, as is expected in (\ref{O7D7}). 
However, they are not the exact discriminant loci but the only approximate values
of the 7-brane positions up to $O(\epsilon^3)$. In particular, $\texttt b_2=0$ 
is not a double zero of the discriminant if we take the $O(\epsilon^3)$
corrections into account.

The Fig.\ref{I0starHiggsbundlelarge} shows a 
closer view of the region around 
$\texttt b_2=0$. There we can see that the two 7-branes split and form 
two clusters with an $f$-plane and $g$-plane.
If we look more closely at one of the clusters, we can see that 
the 7-brane sits inside a small area surrounded by $S$-walls. 

%

This is exactly what we have observed in \cite{dessinonthebase}, 
Fig.6, where we have also considered a deformation of 
a $I_0^*$ fiber in which four D-branes come close together 
from the rest around the origin.  In that case, each of the 
remaining two D-branes forms a cluster with an $f$-plane and 
$g$-plane; they are a {\bf B}-brane and a {\bf C}-brane, 
which are regarded as constituent elements of an O7-plane.

\section{Conclusions}
In this paper 
we have provided three examples of the use of 
the new method of describing the non-localness of 
7-branes by drawing a ``dessin''.  
The first example was a deformation of the $I_0^*$ Kodaira 
fiber. Using the dessin, we were able to immediately 
recognize which pairs of 7-branes 
were (non-)local and compute their monodromies.  
We next solved the mass 
geodesic equation to examine whether the Hanany-Witten effect 
occurred in the example.
Finally, 
we considered the orientifold limit 
in the spectral cover/Higgs bundle approach 
to observe the characteristic configuration 
of an O-plane found previously.

A dessin on the base can display which region is strong- or 
weak-coupling.  It shows how the base 
is covered with such cell regions, which form something like 
a mosaic in F-theory. 
In a perturbative string theory, the expectation value 
of a dilaton is the coupling ``constant'', being an expansion 
parameter in the genus expansion. In a warped compactification 
like a Randall-Sundrum type model or an AdS/CFT set-up, 
there are usually only two regions in which the coupling constant 
is weak or strong. Thus, such a mosaic structure of the 
coupling constant on the base is very characteristic to F-theory. 
Moreover, the region encircled by S-walls is a connected region 
in which one can definitely say that the coupling is weak or strong, 
but which it is is depending on the $SL(2,\ZZ)$ frame, i.e. 
the choice of the fundamental region taken there. 
A dessin on the base can visualize these special features 
of F-theory in a clear manner.

The dessin we considered is a map drawn on the base, which 
is locally homeomorphic to the upper half plane. This is 
thanks to the special Belyi function, which happened to be 
defined by the formula of obtaining the modulus $\tau$ 
via the inverse of the $J$-function. 
This allows us to know which of the 7-branes are  
non-local, easily compute the 
monodromies, and find out whether or not the 
Hanany-Witten effect has occurred, 
as we have demonstrated in this paper. 
We have also found in an orientifold limit 
in the Higgs bundle approach 
the characteristic configuration of an O-plane 
found previously.

We hope that the present analysis will help 
understanding, from a differential geometrical point of view, 
the results obtained in the algebrogeometric/gauge theoretic approaches,
such as the structures of higher codimension and/or higher-rank enhanced 
singularities (e,g, \cite{HKTW,MorrisonTaylor,BoxGraphs,
FtheoryFamilyUnification,MizoguchiTaniAnomaly}).

\vskip 5mm 
We thank the referee of the journal of our previous paper \cite{dessinonthebase}
 for bringing the notion of dessin d'enfant to our attention. 
We also thank S.~Iso, H.~Otsuka, K.~Sakai and T.~Tani for discussions.

%
%

\end{document}